\documentclass{article}

\usepackage{arxiv}

\usepackage{cite}
\usepackage{amsmath,amssymb,amsfonts}
\usepackage{algorithm}
\usepackage{algorithmic}
\usepackage{graphicx}
\usepackage{textcomp}
\usepackage[table]{xcolor}
\usepackage{booktabs}
\usepackage{tabularx}
\usepackage{array}
\usepackage{ragged2e}
\usepackage{caption}
\usepackage{url}
\usepackage{tikz}
\usetikzlibrary{arrows.meta,positioning,shapes.geometric,calc,backgrounds,fit}
\usepackage{pgfplots}
\pgfplotsset{compat=1.18}
\usepackage[hidelinks]{hyperref}

\newcolumntype{L}{>{\RaggedRight\arraybackslash}X}

\def\BibTeX{{\rm B\kern-.05em{\sc i\kern-.025em b}\kern-.08em
    T\kern-.1667em\lower.7ex\hbox{E}\kern-.125emX}}

\definecolor{tableheader}{HTML}{1C355E}
\definecolor{rowtint}{HTML}{EEF2F8}
\definecolor{boxtint}{HTML}{E8EEF7}
\definecolor{gatetint}{HTML}{FBE9D6}

\title{Educating the Agentic Engineer: Curricula, Collaboration, and Continuous Learning in the AI Era}

\author{
 Mamdouh Alenezi \\
  Saudi Data and Artificial Intelligence (SDAIA)\\
  Riyadh, Saudi Arabia
}

\begin{document}
\maketitle
\begin{abstract}
Generative and agentic artificial intelligence (AI) are reconfiguring software and systems engineering from a discipline centered on the human authorship of artifacts toward one centered on the direction, verification, and governance of autonomous systems. This transition demands a new professional archetype---the \emph{agentic engineer}---whose durable value lies in intent specification, orchestration of multi-agent workflows, critical evaluation of machine-generated outputs, and accountable ethical judgment. This article presents an integrative conceptual synthesis of research spanning engineering education, computing education, human--AI interaction, human factors, and the learning sciences to derive an evidence-grounded educational architecture for this archetype. We introduce the ACCEL framework (Agentic Competencies through Curricula, Collaboration, and Enduring Learning), which organizes five competency pillars---intent specification and problem framing; orchestration and delegation; verification, validation, and critical evaluation; ethical governance and accountability; and adaptive self-directed learning---and maps them onto three institutional delivery vectors: curricula, collaboration structures, and continuous learning pathways. Drawing on agency theory, trust-in-automation research, and empirical studies of AI-assisted programming---including randomized and observational evidence that AI benefits are unevenly realized and systematically misperceived---we specify a scaffolded curricular progression, a delegation--verification pedagogical loop for human--AI teaming, redesigned assessment regimes, and governance-literate ethics integration, and we position the framework against current curricular guidelines and international AI-competency frameworks. We articulate the risks the framework is designed to mitigate---automation bias, deskilling, superficial engagement, and diffuse accountability---and close with a research agenda and institutional implications. The analysis argues that educating the agentic engineer requires systemic transformation rather than incremental curricular addition: the unit of instruction must shift from the production of artifacts to the exercise of judgment over increasingly autonomous socio-technical systems.
\end{abstract}


\section{Introduction}
\label{sec:intro}

For most of its modern history, engineering education has rested on a stable implicit contract: universities teach students to \emph{produce} technical artifacts---code, designs, analyses---and professional value accrues to those who produce them well. The rapid maturation of generative and agentic AI systems has destabilized this contract. Contemporary AI agents plan multi-step workflows, write and refactor code, generate architectural alternatives, run analyses, and iterate on their own outputs with diminishing human intervention \cite{alenezi2026delegation, akour2025}; systematic reviews now document large language models and LLM-based agents operating across every phase of the software lifecycle, from requirements elicitation through design, implementation, testing, repair, and maintenance \cite{hou2024, liu2026}. Field and laboratory evidence indicates that these systems can deliver substantial productivity gains on routine knowledge work \cite{peng2023, brynjolfsson2025, dellacqua2023}, while simultaneously introducing new failure modes---plausible-but-wrong outputs, opaque reasoning, and uneven reliability across task types---that demand vigilant human oversight \cite{vaithilingam2022, sarkar2022}. The gains, however, are neither automatic nor universally realized. In a large field deployment, generative AI raised customer-support productivity by 14\% on average but concentrated its benefits among less-experienced workers, acting as a skill-leveler rather than a uniform multiplier \cite{brynjolfsson2025}. More starkly, a randomized controlled trial with experienced open-source developers found that early-2025 AI tooling \emph{slowed} task completion by 19\% even as participants believed it had accelerated them by 20\% \cite{metr2025}---a misperception consistent with large-scale observational evidence that developers' \emph{felt} productivity with AI assistants tracks suggestion-acceptance behavior rather than measured throughput \cite{ziegler2024}. That divergence between perceived and measured benefit is precisely a failure of calibrated judgment---and it foreshadows the educational argument of this article.

The consequence for the profession is a shift in the locus of engineering value: from the volume and speed of manual production toward the quality of \emph{judgment} exercised over autonomous production \cite{alenezi2026rise, alenezi2026collaboration}. We term the professional archetype adequate to this shift the \emph{agentic engineer}: an engineer who specifies intent precisely, decomposes problems for delegation to human and machine collaborators, orchestrates and supervises multi-agent workflows, verifies and validates machine-generated artifacts with the rigor applied to human colleagues' work, and remains ethically and legally accountable for outcomes. The term is deliberately double-voiced. It refers, first, to engineers who work \emph{with} AI agents; and second, in \cite{bandura2006}'s \cite{bandura2006} sense of human agency, to engineers who exercise intentionality, forethought, self-reactiveness, and self-reflectiveness rather than being passively carried by technological change.

Educational institutions are not yet organized to produce this archetype. Curricula remain heavily weighted toward individual artifact production; assessment regimes largely measure unassisted output; ethics is commonly quarantined in standalone modules; and degree structures presume that professional preparation is substantially complete at graduation \cite{sengul2024, johri2023, borenstein2021}. Even the most recent curricular guidelines, which elevate AI to a core knowledge area and adopt competency-based framing \cite{cs2023}, and the international AI-competency frameworks issued for students and teachers \cite{unesco2024students, unesco2024teachers}, stop short of specifying the orchestration, verification, and accountability competencies that supervising \emph{agentic} systems demands. Meanwhile, empirical computing-education research shows that generative AI both disrupts existing pedagogy---undermining traditional assessments and enabling superficial completion of learning tasks---and opens genuinely new pedagogical affordances when integration is deliberate \cite{denny2024, prather2023, becker2023, kazemitabaar2023}. The field therefore faces a design problem: not \emph{whether} to integrate AI into engineering education, but how to re-architect education so that graduates can direct AI rather than be displaced or deskilled by it.

This article addresses that design problem through an integrative conceptual synthesis. Our contributions are fourfold:

\begin{enumerate}
  \item \textbf{Conceptualization.} We provide a theoretically grounded definition of the agentic engineer that couples the socio-cognitive construct of human agency \cite{bandura2006} with empirical findings on AI-native software engineering practice \cite{alenezi2026delegation, bird2023}.
  \item \textbf{Framework.} We introduce ACCEL (\emph{Agentic Competencies through Curricula, Collaboration, and Enduring Learning}), a framework that organizes five competency pillars and maps them onto three institutional delivery vectors (Section~\ref{sec:framework}).
  \item \textbf{Design specification.} We translate the framework into actionable curricular scaffolds, a delegation--verification pedagogical loop for human--AI teaming, assessment redesign principles, and a governance-literate approach to ethics integration (Sections~\ref{sec:curricula}--\ref{sec:assessment}).
  \item \textbf{Research agenda.} We identify open empirical questions concerning deskilling, trust calibration, assessment validity, and institutional change, and state the limitations of a conceptual synthesis (Sections~\ref{sec:discussion}--\ref{sec:limitations}).
\end{enumerate}

The remainder of the article proceeds as follows. Section~\ref{sec:method} describes the synthesis approach. Section~\ref{sec:background} develops the conceptual foundations. Section~\ref{sec:framework} presents the ACCEL framework. Sections~\ref{sec:curricula}--\ref{sec:continuous} elaborate the three delivery vectors, with ethics and assessment treated as cross-cutting threads in Sections~\ref{sec:ethics} and~\ref{sec:assessment}. Section~\ref{sec:discussion} discusses implications and a research agenda; Section~\ref{sec:limitations} states limitations; Section~\ref{sec:conclusion} concludes.

\section{Approach: An Integrative Conceptual Synthesis}
\label{sec:method}

This article is an integrative conceptual synthesis in the tradition of \cite{torraco2005} and \cite{snyder2019}: its purpose is not the exhaustive, protocol-driven coverage of a systematic review, but the construction of a new conceptual architecture from mature yet disconnected literatures. We are explicit about the methodological commitments this entails, and about their limits. The argument was constructed in three stages.

\textbf{Stage 1: Corpus assembly.} Candidate sources were identified through keyword searches of Scopus, Web of Science, the ACM Digital Library, and IEEE Xplore (search window: 1983--2026; core search terms combined \emph{generative AI}, \emph{agentic AI}, \emph{engineering education}, \emph{computing education}, \emph{human--AI collaboration}, \emph{trust in automation}, and \emph{lifelong learning}), supplemented by backward and forward citation chaining from anchor works in each community. From this pool we assembled literature across five research communities whose intersection defines the problem space: (i) engineering education research on curricular change, interdisciplinarity, and lifelong learning \cite{froyd2012, lattuca2017, johri2023}; (ii) computing education research on generative AI in programming instruction \cite{denny2024, prather2023, prather2024gap, finnieansley2022, becker2023, kazemitabaar2023, sengul2024, elnaffar2026}; (iii) human--AI interaction and human factors research on trust, reliance, and automation \cite{amershi2019, lee2004, parasuraman1997, parasuraman2000, bainbridge1983, shneiderman2020, shneiderman2022, vasconcelos2023}; (iv) learning sciences research on active learning, cognitive load, self-regulation, and reflective practice \cite{nasem2018, borte2023, sweller2019, zimmerman2002, schon1983, luckin2019, kasneci2023}; and (v) emerging scholarship and field evidence on AI-native and agentic software engineering \cite{alenezi2026delegation, alenezi2026rise, alenezi2026collaboration, akour2025, bird2023, barke2023, hou2024, liu2026, park2023, dellacqua2023, brynjolfsson2025, ziegler2024, metr2025}. Where the argument required theoretical machinery for delegation itself, we drew additionally on agency theory in its canonical principal--agent form \cite{jensen1976} and its recent extension to delegation to and from agentic information-system artifacts \cite{baird2021}. Sources were selected for peer-reviewed standing, citational influence, and direct relevance to the education of engineers who supervise autonomous systems; recent preprints and working papers were admitted only where they report emerging practice or field evidence not yet available in the archival literature, and every claim they support in this article is additionally anchored to at least one peer-reviewed source wherever possible.

\textbf{Stage 2: Thematic integration.} Following the logic of integrative conceptual review \cite{torraco2005}, we iteratively coded the corpus for competency claims (what agentic engineers must be able to do), pedagogical claims (how such competencies are developed), and risk claims (what failure modes education must mitigate). Convergent claims across at least two research communities were promoted to framework elements; claims supported within a single community were retained as design suggestions and flagged as such. Coding was performed by the author; the absence of independent coding and inter-rater checks is acknowledged as a limitation in Section~\ref{sec:limitations}, and is partially offset by the convergence rule just stated, which requires cross-community corroboration before any claim enters the framework.

\textbf{Stage 3: Framework construction and stress-testing.} The resulting elements were organized into the ACCEL framework and stress-tested against three adversarial questions: Does each element survive plausible near-term advances in AI capability? Does each element have at least one concrete curricular or assessment instantiation? And does the framework as a whole mitigate---rather than merely acknowledge---the documented risks of automation bias, deskilling, and superficial engagement \cite{parasuraman1997, bainbridge1983, kazemitabaar2023}?

The limitations inherent in this approach---notably the absence of primary empirical validation---are addressed in Section~\ref{sec:limitations}.

\section{Conceptual Foundations}
\label{sec:background}

\subsection{From tools to agents: the changing object of engineering work}
\label{sec:tools-to-agents}

The distinction between an AI \emph{tool} and an AI \emph{agent} is consequential for education. A tool augments a human-executed task: a compiler, a static analyzer, an autocomplete engine. An agent, by contrast, accepts an intent, plans a course of action, executes multi-step workflows---often invoking other tools and agents---and returns candidate outcomes for human review \cite{alenezi2026delegation, akour2025}. The distinction is not merely rhetorical: LLM-based agents combining planning, memory, tool invocation, and iterative self-correction are now documented across the entire software lifecycle, including multi-agent architectures in which specialized planner, coder, tester, and reviewer agents coordinate under human governance \cite{liu2026, park2023}, and systematic review of the broader LLM-for-software-engineering literature confirms the shift of engineering effort from authoring artifacts to prompting, reviewing, and validating machine-generated ones \cite{hou2024}. Classic human-factors theory supplies the vocabulary for reasoning about this shift: \cite{parasuraman2000} model automation as operating at graded levels across four cognitive stages---information acquisition, analysis, decision selection, and action execution---and show that the design question is never \emph{whether} to automate but \emph{which functions, to what level, with what human oversight}. Agentic systems push automation to high levels across all four stages simultaneously, which is exactly the configuration in which their model predicts the greatest supervisory burden.

Studies of AI-assisted development document the transitional phase: developers using code-generation assistants report substantial acceleration on routine tasks alongside recurring difficulties in understanding, debugging, and trusting generated code \cite{vaithilingam2022, bird2023, sarkar2022}. Grounded-theory observation adds interaction-level texture: programmers engage code-generating models in two distinct modes---\emph{acceleration}, in which the programmer already knows the intended code and vets suggestions in seconds against a mental template, and \emph{exploration}, in which the model supplies candidate approaches that demand deliberate prompting, comparison, and validation---and effective use of either mode is a learned regulatory meta-skill with documented expert--novice differences, not a byproduct of language knowledge \cite{barke2023}. Controlled evidence quantifies both edges of the capability boundary: large productivity gains on tasks within the technology's competence frontier, and \emph{degraded} performance when workers rely on AI beyond that frontier \cite{peng2023, dellacqua2023}. \cite{dellacqua2023} memorably describe this uneven boundary as a ``jagged frontier,'' and field evidence shows the realized benefit is further moderated by worker experience, with the largest gains accruing to novices on routine work \cite{brynjolfsson2025, peng2023}. Subsequent randomized evidence sharpens the point: experienced developers using early-2025 tools on mature, high-context codebases were measurably slower with AI assistance yet perceived themselves faster, a roughly forty-percentage-point gap between forecast and measured effect \cite{metr2025}. The contrast with the 56\% speedup observed on bounded, well-specified greenfield tasks \cite{peng2023} is instructive: the same class of tool produces opposite effects in different task regimes, with contextual complexity, quality standards, and verification cost as the moderating variables. Self-perception, meanwhile, is an unreliable instrument for locating the frontier---perceived productivity tracks suggestion-acceptance rates rather than measured output \cite{ziegler2024}---and the central educational implication of the present article is that mapping, probing, and respecting that frontier is a learnable, teachable professional competency rather than a matter of intuition.

As autonomy increases, the engineer's relationship to the system shifts from operator to supervisor and from author to editor-of-record. Classic human factors research anticipated the hazards of exactly this shift: \cite{bainbridge1983} showed that automating routine work paradoxically \emph{raises} the skill demands on the human supervisor, who must intervene precisely when the automation fails in unfamiliar ways; \cite{parasuraman1997} catalogued the misuse (over-reliance) and disuse (under-reliance) patterns that follow poorly calibrated trust; and \cite{lee2004} demonstrated that appropriate reliance depends on the human's ability to assess automation performance across contexts. Engineering education has largely ignored this literature because its graduates were artifact producers, not automation supervisors. That exemption has ended.

\subsection{Agency as the organizing construct}
\label{sec:agency}

We ground the agentic engineer in \cite{bandura2006}'s socio-cognitive theory of human agency, which identifies four core properties: \emph{intentionality} (forming action plans and strategies), \emph{forethought} (anticipating outcomes and setting goals), \emph{self-reactiveness} (motivating and regulating execution), and \emph{self-reflectiveness} (examining one's own functioning and correcting course). Each property maps directly onto a demand of AI-native practice: intentionality onto precise intent specification and problem framing; forethought onto anticipating agent failure modes and designing guardrails before delegation; self-reactiveness onto monitoring, intervening in, and steering agentic workflows in flight; and self-reflectiveness onto post-hoc evaluation of both the artifact and the human--AI process that produced it, including one's own reliance behavior.

The socio-cognitive account is complemented by a second theoretical lens: agency theory in its principal--agent form. The canonical analysis of delegation identifies its enduring costs---information asymmetry between principal and agent, divergence between delegated intent and executed behavior, and the monitoring expenditure required to contain both \cite{jensen1976}. \cite{baird2021} extend this apparatus to information systems that are themselves agentic, arguing that as artifacts acquire autonomy the research question shifts from technology \emph{use} to bidirectional \emph{delegation}: humans delegate tasks to agentic artifacts, artifacts delegate sub-decisions and exceptions back to humans, and the appropriateness of each delegation depends on task characteristics, artifact capability, and the allocation of accountability. Transposed to engineering education, this framing converts orchestration from a vague soft skill into a structured decision problem---what to delegate, under what monitoring regime, with what reversibility---whose costs and failure modes are theoretically characterized in advance.

Together the two lenses do theoretical work that a purely skills-based account cannot. They explain why the agentic engineer is not reducible to a checklist of tool proficiencies: agency is a disposition exercised over whatever tools exist, which is precisely what makes it durable across technology generations, and delegation is a principal's problem that persists regardless of which agent technology occupies the other side of the contract. The grounding also connects the professional construct to well-developed educational machinery---self-regulated learning \cite{zimmerman2002}, reflective practice \cite{schon1983}, and active learning \cite{borte2023}---giving curriculum designers established levers rather than requiring pedagogical invention from scratch.

\subsection{What the empirical record already shows}
\label{sec:empirical}

Four findings from the recent empirical record constrain any credible educational response.

First, \textbf{generative AI collapses traditional novice tasks}. Code-generation models solve typical introductory programming assessments at or above student level \cite{finnieansley2022}, undermining the signaling value of unassisted-production assessment and forcing a re-examination of what introductory courses are \emph{for} \cite{becker2023, denny2024}.

Second, \textbf{unstructured access produces divergent learning outcomes}. Novices given AI code generators complete tasks faster without uniform learning loss, but exhibit distinct usage profiles: some use generation as scaffolding for understanding, while others develop dependency patterns associated with weaker subsequent unassisted performance \cite{kazemitabaar2023}. Observational replication work confirms and extends the divide: generative AI accelerates already-capable novices while compounding the metacognitive difficulties of struggling ones, who frequently finish with an ``illusion of competence''---believing they performed better than they did \cite{prather2024gap}. A recent systematic review of 58 studies of AI agents in programming education reports over-reliance producing superficial learning in roughly two-thirds of the studies examined \cite{elnaffar2026}. The workplace analogue is now documented at scale: AI assistance functions as a skill-leveler that lifts weaker performers most \cite{brynjolfsson2025, peng2023}---an equity opportunity, but also a mechanism by which novices can experience an artificial inflation of capability precisely when their independent verification skill is weakest. Instructional structure, not access per se, determines whether AI assistance builds or borrows competence---an instance of the general finding that AI interventions unmoored from cognitive theory and coherent instructional design underperform \cite{luckin2019, sweller2019, zawacki2019, kasneci2023}.

Third, \textbf{intent specification is a distinct, isolable bottleneck}. A causal-intervention study of beginning students' failed text-to-code prompts tested whether the deficit lies in \emph{style} (lacking technical vocabulary) or \emph{substance} (not understanding how much information the model needs): surgically replacing informal vocabulary with correct terminology largely failed to rescue performance, while the information content of the prompt---whether it specified inputs, outputs, edge cases, and behavioral constraints---predicted success \cite{lucchetti2025}. The same study documents a characteristic ``stuck'' revision pattern in which students respond to failed generations with cosmetic rewording rather than added specification, iterating in place without converging. Specification competence is thus conceptual rather than lexical; it will not be produced by tool exposure, prompt templates, or vocabulary drills, and it must be taught and assessed in its own right.

Fourth, \textbf{professional practice is reorganizing around review and orchestration}. Studies of AI-assisted developers show effort shifting from writing code to specifying, prompting, evaluating, and integrating it \cite{bird2023, sarkar2022, barke2023}; systematic reviews document LLMs and LLM-based agents deployed across requirements, design, implementation, testing, debugging, repair, and maintenance \cite{hou2024, liu2026}; and accounts of agentic DevOps document pipelines in which agents propose, test, and deploy changes under human governance \cite{akour2025, alenezi2026collaboration}. Notably, the open challenges catalogued by the agentic-SE research community read as an inverted competency specification for its human supervisors: agents remain unreliable at requirement disambiguation, their end-to-end autonomy amplifies error propagation across pipeline stages, and orchestration quality often dominates raw model capability in determining outcomes \cite{liu2026}. Education that continues to optimize for unassisted production is therefore optimizing for a shrinking slice of professional activity.

Together these findings define the design constraints: preserve foundational understanding (because supervision without comprehension is vacuous), teach calibrated reliance (because the frontier is jagged), teach specification explicitly (because underspecification, not phrasing, is the documented novice failure), restructure assessment (because unassisted production no longer discriminates), and institutionalize continuous adaptation (because the frontier moves).

\section{The ACCEL Framework}
\label{sec:framework}

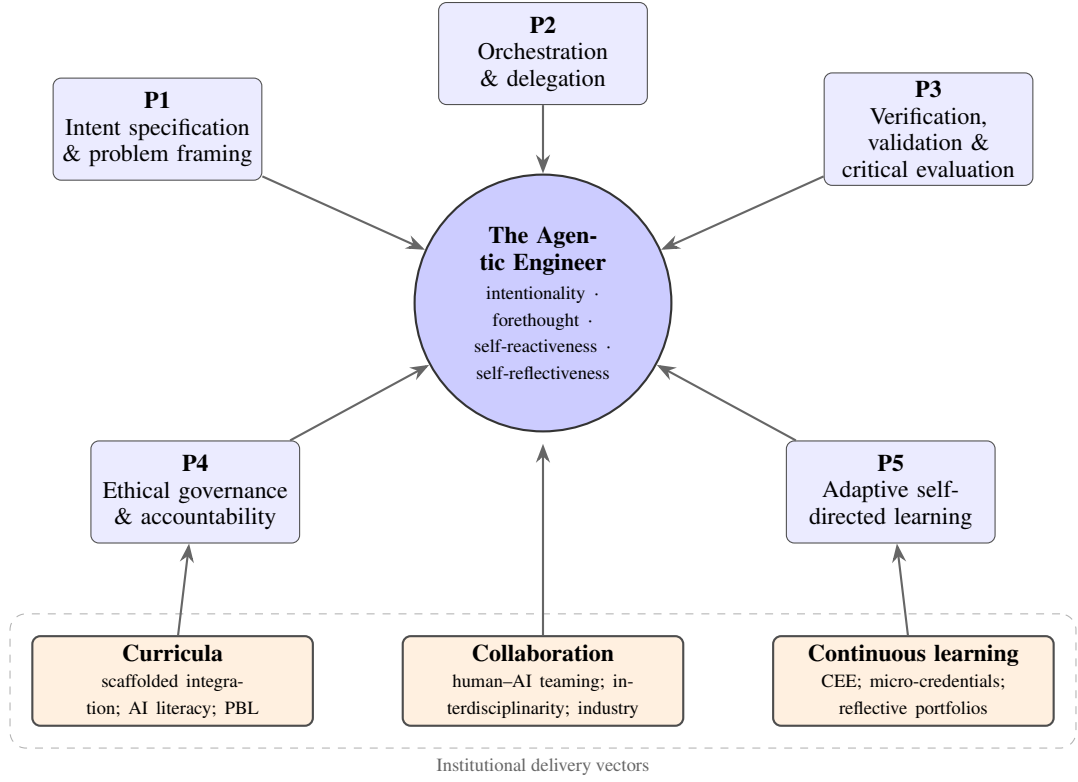
\begin{figure}[t]
\centering
\begin{tikzpicture}[
  font=\small,
  pillar/.style={rectangle, rounded corners=3pt, draw=black!70, fill=blue!8,
                 text width=2.55cm, minimum height=1.35cm, align=center, inner sep=3pt},
  core/.style={circle, draw=black!80, thick, fill=blue!20, text width=2.5cm,
               align=center, inner sep=2pt},
  vector/.style={rectangle, rounded corners=3pt, draw=black!70, thick, fill=orange!12,
                 text width=3.5cm, minimum height=1.1cm, align=center, inner sep=3pt},
  arr/.style={-{Stealth[length=2.5mm]}, thick, black!60}
]
\node[core] (core) at (0,0) {\textbf{The Agentic Engineer}\\[1pt]\scriptsize intentionality \(\cdot\) forethought \(\cdot\) self-reactiveness \(\cdot\) self-reflectiveness};

\node[pillar] (p1) at (-5.1, 2.3) {\textbf{P1}\\Intent specification \& problem framing};
\node[pillar] (p2) at (0, 3.3)    {\textbf{P2}\\Orchestration \& delegation};
\node[pillar] (p3) at (5.1, 2.3)  {\textbf{P3}\\Verification, validation \& critical evaluation};
\node[pillar] (p4) at (-4.6, -2.5){\textbf{P4}\\Ethical governance \& accountability};
\node[pillar] (p5) at (4.6, -2.5) {\textbf{P5}\\Adaptive self-directed learning};

\draw[arr] (p1) -- (core);
\draw[arr] (p2) -- (core);
\draw[arr] (p3) -- (core);
\draw[arr] (p4) -- (core);
\draw[arr] (p5) -- (core);

\node[vector] (v1) at (-4.9, -5.0) {\textbf{Curricula}\\\scriptsize scaffolded integration; AI literacy; PBL};
\node[vector] (v2) at (0, -5.0)    {\textbf{Collaboration}\\\scriptsize human--AI teaming; interdisciplinarity; industry};
\node[vector] (v3) at (4.9, -5.0)  {\textbf{Continuous learning}\\\scriptsize CEE; micro-credentials; reflective portfolios};

\draw[arr] (v1) -- (p4);
\draw[arr] (v2) -- ($(core.south)+(0,-0.15)$);
\draw[arr] (v3) -- (p5);

\begin{scope}[on background layer]
\node[rectangle, rounded corners=6pt, draw=black!30, dashed,
      fit=(v1)(v2)(v3), inner sep=8pt, label={[black!60]below:\scriptsize Institutional delivery vectors}]
      (band) {};
\end{scope}
\end{tikzpicture}
\caption{The ACCEL framework. Five competency pillars (P1--P5) converge on the agentic engineer, whose core is defined by Bandura's four properties of human agency. Three institutional delivery vectors---curricula, collaboration structures, and continuous learning pathways---develop and sustain the pillars across the professional lifespan.}
\label{fig:accel}
\end{figure}

We now integrate these foundations into a single architecture. ACCEL---\emph{Agentic Competencies through Curricula, Collaboration, and Enduring Learning}---comprises (i) a competency core of five pillars that specify \emph{what} the agentic engineer must be able to do, and (ii) three institutional delivery vectors that specify \emph{how} educational systems develop those pillars. Figure~\ref{fig:accel} depicts the architecture; Table~\ref{tab:pillars} operationalizes the pillars as learning outcomes and assessment evidence.

\subsection{The five competency pillars}
\label{sec:pillars}

\textbf{P1: Intent specification and problem framing.} The upstream competency on which all delegation depends. It encompasses eliciting and formalizing requirements, decomposing ill-structured problems into delegable units, and expressing intent as executable specifications with explicit constraints, acceptance criteria, and guardrails \cite{alenezi2026delegation}. Prompt engineering is the currently visible surface of this pillar, but the durable competency is specification discipline---a lineage engineering education already possesses through requirements engineering \cite{se2014} and which now requires generalization to machine addressees. The pillar's status as a genuine, isolable skill is empirically established: novice failure at AI-mediated code generation is driven by underspecification of the machine interlocutor's information requirements, not by deficient phrasing, and it persists under vocabulary correction \cite{lucchetti2025}. Requirement disambiguation is likewise the documented weak point of current agentic systems themselves \cite{liu2026}, making human specification competence the binding input to hybrid-team performance.

\textbf{P2: Orchestration and delegation.} The capacity to allocate work across a hybrid team of humans and agents: selecting appropriate agents for sub-tasks, sequencing and parallelizing workflows, managing hand-offs and shared context, and adapting the allocation as evidence about agent performance accumulates \cite{akour2025, alenezi2026collaboration}. The pillar is theoretically structured by delegation theory---what to delegate, under what monitoring regime, with what accountability allocation \cite{jensen1976, baird2021}---and by the levels-of-automation model, which frames each delegation as a choice of automation level per cognitive stage rather than a binary hand-off \cite{parasuraman2000}. Its technical substrate is now well documented: multi-agent architectures coordinate specialized planner, coder, tester, and reviewer agents through structured communication, and their emergent interactions can be difficult to predict \cite{liu2026, park2023}, which is precisely why orchestration quality, not raw model capability, often dominates outcomes. This pillar recasts classical project management and software architecture skills for teams whose members include non-human contributors of variable and shifting reliability.

\textbf{P3: Verification, validation, and critical evaluation.} The defining supervisory competency: reviewing machine-generated artifacts with the rigor applied to human colleagues' work, detecting plausible-but-wrong outputs, designing test regimes and acceptance gates for AI contributions, interpreting uncertainty, and calibrating reliance to demonstrated performance \cite{lee2004, vaithilingam2022}. Empirically, this is where AI-assisted practice most often fails \cite{dellacqua2023}; pedagogically, it is where automation bias and deskilling must be confronted directly \cite{parasuraman1997, bainbridge1983}. Importantly, overreliance is not immutable: experimental work shows it responds to the cost--benefit structure of engagement, declining when explanations make verification cheap relative to blind acceptance and when task stakes make errors salient \cite{vasconcelos2023}---which converts critical evaluation from an exhortation into a designable property of learning environments.

\textbf{P4: Ethical governance and accountability.} Beyond conventional professional ethics, this pillar comprises literacy in algorithmic bias and fairness \cite{memarian2023}, transparency and interpretability demands for safety-relevant systems, intellectual-property and privacy obligations attached to generated artifacts, and working knowledge of the maturing regulatory apparatus---risk-management frameworks and statutory regimes---within which engineered AI systems must now operate \cite{nist2023, euaiact2024, borenstein2021, unesco2021}. The pillar's design rationale follows \cite{mittelstadt2019}: high-level ethical principles cannot by themselves guarantee ethical AI, because they lack the technical and institutional mechanisms of implementation; the engineer must therefore be able to \emph{translate} principles into system constraints---traceable decision paths, audit trails, human-oversight points, and fail-safes---in the spirit of human-centered AI design, which treats reliability, safety, and trustworthiness as engineered properties supporting human control rather than aspirations appended to it \cite{shneiderman2020, shneiderman2022}. The pillar's core commitment is non-delegable accountability: authority over decisions may be shared with machines; responsibility may not.

\textbf{P5: Adaptive self-directed learning.} The meta-competency that keeps the other four current as the capability frontier moves. It comprises self-regulated learning strategies \cite{zimmerman2002}, reflective practice on one's own human--AI process \cite{schon1983}, systematic evaluation of unfamiliar tools and models, and construction of a personal learning infrastructure---communities of practice, curated information flows, credentialing pathways---that sustains professional currency across a career \cite{fischer2000}. The pillar rests on the consensus synthesis of the learning sciences: durable expertise develops through active engagement, deliberate practice, feedback, and metacognitive monitoring of one's own understanding, and self-regulation is a trainable capability rather than a fixed trait \cite{nasem2018}.

\begin{table}[t]
\centering
\caption{The ACCEL competency pillars operationalized as exemplar learning outcomes and assessment evidence.}
\label{tab:pillars}
\small
\begin{tabularx}{\textwidth}{@{}p{2.7cm} L L@{}}
\toprule
\textbf{Pillar} & \textbf{Exemplar learning outcomes} & \textbf{Exemplar assessment evidence} \\
\midrule
P1 Intent specification \& problem framing &
Decompose an ill-structured problem into delegable units; author executable specifications with acceptance criteria and guardrails; justify framing choices against stakeholder needs. &
Specification documents graded for completeness and testability; comparative critique of alternative decompositions; traceability from requirements to delegated tasks. \\
\addlinespace
P2 Orchestration \& delegation &
Allocate sub-tasks across humans and agents with explicit rationale; design multi-agent workflows with hand-off and context-management protocols; re-plan when agent performance deviates. &
Orchestration logs and workflow diagrams; design reviews defending allocation decisions; documented re-planning episodes in project retrospectives. \\
\addlinespace
P3 Verification, validation \& critical evaluation &
Detect seeded defects in AI-generated artifacts; design acceptance gates and test regimes for machine contributions; demonstrate calibrated reliance across task types. &
Defect-detection exercises with known ground truth; review reports on AI contributions; reliance-calibration records comparing trust decisions with measured agent accuracy. \\
\addlinespace
P4 Ethical governance \& accountability &
Conduct bias and impact assessments; map system decisions to accountable humans; apply risk-management and regulatory requirements to a concrete deployment. &
Impact-assessment artifacts; accountability matrices; red-team exercise reports with reflective debriefs. \\
\addlinespace
P5 Adaptive self-directed learning &
Systematically evaluate an unfamiliar AI tool against task requirements; maintain a reflective record of one's own human--AI process; construct and defend a personal development plan. &
Tool-evaluation reports with explicit criteria; reflective portfolios; evidence of self-initiated learning (micro-credentials, community contributions). \\
\bottomrule
\end{tabularx}
\end{table}

\subsection{The three delivery vectors}

The pillars are developed through three mutually reinforcing institutional vectors. \emph{Curricula} (Section~\ref{sec:curricula}) provide vertically and horizontally integrated, scaffolded development of the pillars across a degree program. \emph{Collaboration structures} (Section~\ref{sec:collaboration}) supply the authentic hybrid-team contexts---human--AI, interdisciplinary, and industry-connected---in which the pillars are exercised under realistic conditions. \emph{Continuous learning pathways} (Section~\ref{sec:continuous}) extend development beyond graduation through continuing engineering education, micro-credentials, and reflective professional practice. Ethics (Section~\ref{sec:ethics}) and assessment (Section~\ref{sec:assessment}) are deliberately treated as cross-cutting threads woven through all three vectors rather than as vectors of their own: the framework's central design claim is that quarantining either one recreates precisely the fragmentation it seeks to repair.

\section{Vector I --- Curricula: Scaffolded Integration Rather Than Additive Modules}
\label{sec:curricula}

\subsection{Integration, not addition}

The dominant institutional reflex---adding an ``AI tools'' elective to an otherwise unchanged program---fails on both theoretical and empirical grounds. Isolated, tool-specific instruction does not produce transferable competence \cite{memarian2023, zawacki2019}, and bolting AI onto unrevised pedagogy erects barriers to the active learning on which durable understanding depends \cite{borte2023}. The historical analysis of \cite{froyd2012} shows that engineering education's major advances came from systemic shifts---toward outcomes-based accreditation, learning-sciences-informed pedagogy, and context-rich instruction---not from course-level additions. ACCEL therefore prescribes \emph{vertical} integration (pillar development sequenced across program years) and \emph{horizontal} integration (pillar exercise embedded within disciplinary courses, from mechanics to databases), with AI literacy in \cite{long2020}'s sense---knowing what AI is, what it can and cannot do, and how to critique its outputs---treated as a program-wide foundation rather than a course topic, consistent with its recent institutional codification in international competency frameworks \cite{unesco2024students} and in the CS2023 guidelines' elevation of AI to a core, cross-cutting knowledge area \cite{cs2023}.

\subsection{A scaffolded progression}

Table~\ref{tab:scaffold} specifies a four-stage progression, each stage defined by the autonomy granted to AI collaborators and the corresponding supervisory demand placed on the student. The progression operationalizes two learning-sciences constraints. First, \emph{foundations precede supervision}: students cannot verify what they cannot understand, so early stages deliberately restrict AI assistance in foundational skill-building while using AI tutoring to support---not substitute for---practice \cite{sweller2019, kazemitabaar2023}; this ordering follows directly from the consensus finding that expertise is built through active engagement and deliberate practice rather than exposure to solutions \cite{nasem2018}. Second, \emph{difficulty is granted, not evaded}: at each stage, AI autonomy increases only as verification competence is demonstrated, preventing the dependency profiles observed under unstructured access \cite{kazemitabaar2023, becker2023}. The staging also responds to the broader analysis of LLMs in education: the technologies shift learners from producers of every artifact toward supervisors and evaluators of machine-generated work, and educational objectives must move correspondingly toward problem formulation, verification, judgment, and ethical reasoning \cite{kasneci2023}.

\begin{table}[t]
\centering
\caption{Scaffolded curricular progression for agentic engineering education. AI autonomy is expanded stage-by-stage, contingent on demonstrated verification competence.}
\label{tab:scaffold}
\small
\begin{tabularx}{\textwidth}{@{}p{3.7cm} p{4.2cm} L L@{}}
\toprule
\textbf{Stage} & \textbf{AI role} & \textbf{Student focus} & \textbf{Dominant pillars} \\
\midrule
1. Foundations (Year 1) &
Tutor and explainer; generation restricted in assessed work &
Core disciplinary concepts; unassisted problem solving; AI literacy; first critique of AI outputs against known ground truth &
P3 (nascent), P5 \\
\addlinespace
2. Assisted practice (Year 2) &
Pair collaborator on bounded tasks &
Specification writing; systematic review of AI contributions; seeded-defect detection; documentation of AI usage &
P1, P3 \\
\addlinespace
3. Supervised delegation (Year 3) &
Semi-autonomous agent on multi-step tasks &
Workflow orchestration; acceptance gates; reliance calibration; interdisciplinary team projects with AI members &
P2, P3, P4 \\
\addlinespace
4. Orchestration (Year 4 / capstone) &
Multi-agent teams across the full lifecycle &
End-to-end intent-to-outcome responsibility; governance artifacts; reflective analysis of the human--AI process &
All pillars \\
\bottomrule
\end{tabularx}
\end{table}

\subsection{Project-based learning as the load-bearing pedagogy}

Project-based learning (PBL) is the natural vehicle for the upper stages because agentic competence is exercised, not recited: it emerges from consequential decisions about delegation, trust, and integration under ambiguity \cite{chen2021}, and the engagement dividends of well-designed experiential formats in software engineering education are durable rather than novelty effects \cite{akour2024}. In an ACCEL-aligned capstone, students define problem boundaries, select and configure AI collaborators, negotiate evolving goals, and remain accountable for the delivered system. Critically, assessment weight shifts from the final artifact---which agents can increasingly produce---to the orchestration record: the quality of specifications, the defensibility of delegation decisions, the rigor of verification, and the honesty of reflection \cite{denny2024, prather2023}. Section~\ref{sec:assessment} develops this shift in full.

\subsection{Adaptive platforms under pedagogical control}

AI-powered adaptive learning platforms can personalize pacing, diagnose misconceptions, and simulate complex engineering scenarios at scale. Their integration, however, must be governed by the finding that educational technology unanchored in learning theory prioritizes novelty over effectiveness \cite{luckin2019, zawacki2019}: adaptive difficulty must respect cognitive-load constraints \cite{sweller2019}, tutoring must scaffold rather than answer, and educators must retain design authority over the systems that mediate their students' learning. The agentic-engineering program is itself an appropriate site for this scrutiny---students who dissect their own adaptive platform's recommendation logic are simultaneously exercising P3 and P4 on a system with immediate personal stakes.

\section{Vector II --- Collaboration: Hybrid Teams as the Unit of Practice}
\label{sec:collaboration}

\subsection{The delegation--verification loop}

\begin{figure}[t]
\centering
\begin{tikzpicture}[
  font=\small,
  stage/.style={rectangle, rounded corners=3pt, draw=black!70, fill=blue!8,
                text width=2.7cm, minimum height=1.05cm, align=center, inner sep=3pt},
  agent/.style={rectangle, rounded corners=3pt, draw=black!70, fill=orange!15,
                text width=2.7cm, minimum height=1.05cm, align=center, inner sep=3pt},
  arr/.style={-{Stealth[length=2.5mm]}, thick, black!65},
  lab/.style={black!60, font=\scriptsize}
]
\node[stage] (s1) at (0,0)      {\textbf{1. Specify intent}\\\scriptsize frame, decompose, set acceptance criteria (P1)};
\node[stage] (s2) at (4.6,0)    {\textbf{2. Delegate}\\\scriptsize select agents, set guardrails (P2, P4)};
\node[agent] (s3) at (9.2,0)    {\textbf{3. Agent executes}\\\scriptsize plans, generates, iterates};
\node[stage] (s4) at (9.2,-3.0) {\textbf{4. Verify \& validate}\\\scriptsize test, inspect, calibrate reliance (P3)};
\node[stage] (s5) at (4.6,-3.0) {\textbf{5. Integrate or re-delegate}\\\scriptsize accept, revise intent, or reassign (P2)};
\node[stage] (s6) at (0,-3.0)   {\textbf{6. Reflect \& capture rationale}\\\scriptsize document decisions, update trust model (P5)};

\draw[arr] (s1) -- (s2);
\draw[arr] (s2) -- (s3);
\draw[arr] (s3) -- (s4);
\draw[arr] (s4) -- (s5);
\draw[arr] (s5) -- (s6);
\draw[arr] (s6.west) -- ++(-0.7,0) |- (s1.west);
\draw[arr, dashed] (s5.north) -- (s2.south)
     node[midway, right, lab] {re-delegation};
\draw[arr, dashed] (s4.north) -- (s3.south)
     node[midway, right, lab] {rejection};
\end{tikzpicture}
\caption{The delegation--verification loop. Students traverse the loop explicitly, producing assessable artifacts at every stage; accountability checks (P4) gate integration of any irreversible change. Dashed edges mark the rejection and re-delegation paths whose exercise distinguishes calibrated from credulous reliance.}
\label{fig:loop}
\end{figure}
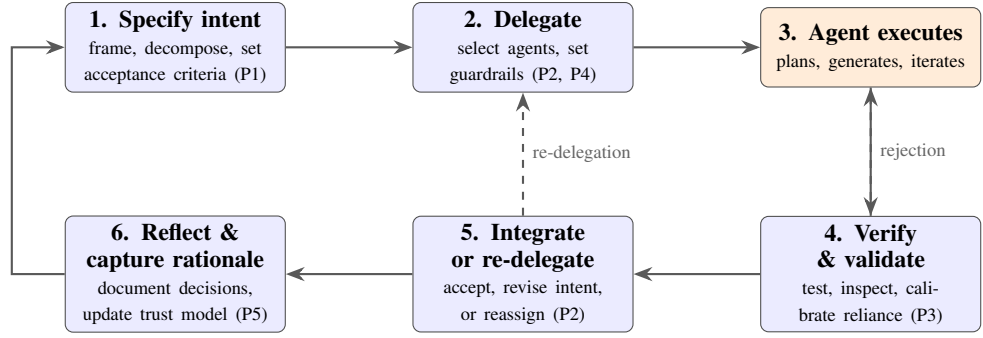

The pedagogical core of human--AI teaming instruction is a disciplined loop that students traverse repeatedly, first slowly and explicitly, later fluently (Figure~\ref{fig:loop}). The loop renders visible---and therefore teachable and assessable---the supervisory cycle that expert AI-native practitioners execute tacitly \cite{bird2023, sarkar2022}. Its stages instantiate the pillars in sequence: intent specification and decomposition (P1), delegation with explicit guardrails (P2), agent execution, verification against acceptance criteria (P3), integration or re-delegation, and reflective rationale capture (P5), with accountability checks (P4) gating irreversible actions.

Three design features guard against the loop degenerating into ritual. First, curricula must engineer encounters with \emph{failure}: exercises seeded with plausible-but-wrong agent outputs, ``AI coworker'' simulations in which over-trust has visible consequences, and post-incident debriefs \cite{parasuraman1997, lee2004}. Students who have never caught an agent being confidently wrong have not learned verification; they have learned deference. Crucially, calibration must be \emph{measured} rather than self-reported: even experienced professionals systematically misjudge whether AI assistance is helping them \cite{metr2025, ziegler2024}, so curricula should confront students with objective records of their own acceptance decisions against ground truth, not with reflection alone. Second, the cost structure of verification must be deliberately engineered. Overreliance is sensitive to the relative effort of engaging critically versus accepting blindly: when explanations and supporting evidence make verification cheap, and when stakes make errors salient, people verify more \cite{vasconcelos2023}. Loop-based exercises should therefore require agents (and students configuring them) to surface reasoning, intermediate artifacts, and confidence signals that lower the cost of inspection---simultaneously training students to \emph{demand} inspectable output as a delegation precondition. Because verification depth should also vary by interaction mode---shallow pattern-matching suffices in acceleration mode, while exploration mode demands documentation lookup, comparison, and testing \cite{barke2023}---students must learn to diagnose which mode they are in and match their evaluation strategy to it. Third, guidance on human--AI interaction---making capability boundaries visible, supporting efficient correction, calibrating expectations---should be taught as design knowledge students apply when they build agentic systems for others, not merely experience as users \cite{amershi2019, shneiderman2020, shneiderman2022}.

\subsection{Interdisciplinary teams and AI as boundary object}

Modern AI-enabled systems entangle computer science, data science, security, human factors, law, ethics, and domain expertise; interdisciplinary competence is accordingly a validated engineering learning outcome with a developed research base \cite{lattuca2017}. Agentic practice sharpens the demand in a specific way: AI agents function as \emph{boundary objects} that silently encode assumptions from multiple fields---statistical assumptions from their training regimes, value assumptions from their alignment procedures, domain assumptions from their data. Interrogating those assumptions is intrinsically a multi-disciplinary act. ACCEL therefore prescribes team-based experiences in which engineering students collaborate with peers from data science, design, and the humanities to audit and adapt agent behavior for domain-specific constraints---rehearsing the distributed cognition of professional practice while exercising P3 and P4 on live systems.

\subsection{Industry partnership as curricular metabolism}

AI capabilities evolve on cycles measured in months; university curriculum revision operates on cycles measured in years. Sustained industry partnership---mentored projects, internships in AI-native teams, challenge-based learning, co-designed learning outcomes---is therefore not enrichment but metabolic necessity, the mechanism by which programs sense and absorb changes at the practice frontier \cite{johri2023, sengul2024}. National capability-building initiatives, in which academies co-locate curriculum design with government and industry demand signals, exemplify the institutional form this metabolism can take \cite{alenezi2026rise}.

\section{Vector III --- Continuous Learning: The Career-Length Program}
\label{sec:continuous}

If the capability frontier moves continuously, terminal degrees cannot be terminal. ACCEL treats the degree as the first phase of a career-length program with three components.

\textbf{Self-directed learning capacity as a designed outcome.} The disposition to identify one's own knowledge gaps, locate resources, and regulate learning is a trainable competency, not a temperament \cite{zimmerman2002, fischer2000}. Programs build it deliberately: learning-to-learn instruction embedded in disciplinary courses, reflective portfolios that make students' own learning processes objects of analysis \cite{schon1983}, and repeated tool-evaluation exercises (P5) that rehearse the professional act of confronting an unfamiliar system and systematically establishing what it can be trusted to do.

\textbf{Structured continuing engineering education.} Reskilling and upskilling pathways---stackable micro-credentials, modular certificates, executive programs---give the post-graduation phase institutional form. Taxonomic standardization of continuing engineering education enables benchmarking and portability across providers and borders, and alignment between credential frameworks and evolving job architectures allows both institutions and governments to forecast and provision workforce capability \cite{alenezi2026rise, unesco2021}. The design principle carried over from Section~\ref{sec:curricula} applies with full force: continuing education must develop pillar competencies, not perishable tool proficiencies.

\textbf{Personal learning infrastructure.} Between formal episodes, professionals sustain currency through infrastructure they curate themselves: communities of practice, monitored research and standards flows, personal experimentation sandboxes, and---reflexively---AI-powered learning tools whose recommendations they evaluate with the same calibrated skepticism they apply to any agent (P3 applied to P5). Graduates who leave with such an infrastructure already operating, seeded by the reflective portfolio and tool-evaluation practices above, cross the education-to-practice boundary without interrupting their learning.

\section{Cross-Cutting Thread I --- Ethics and Governance}
\label{sec:ethics}

Standalone ethics modules demonstrably underperform: codes of conduct imparted in isolation do not transfer to technical decision-making \cite{borenstein2021, memarian2023}. The deeper diagnosis is \cite{mittelstadt2019}'s \cite{mittelstadt2019}: unlike medicine, AI practice lacks the common aims, professional norms, and accountability mechanisms that make principle-based ethics effective, so principles alone cannot guarantee ethical systems---they must be translated into concrete technical and institutional mechanisms. That translation is engineering work, and it belongs in the engineering curriculum. ACCEL therefore embeds ethical reasoning as a continuous thread with three strands.

First, \emph{situated ethical reasoning within technical courses}: bias audits inside machine-learning coursework, privacy analyses inside database and systems design, accountability mapping inside software architecture---so that ethical analysis is encountered as part of engineering judgment, not adjacent to it \cite{memarian2023}.

Second, \emph{adversarial and reflective exercises}: structured red-team activities in which students deliberately probe AI systems for bias, unsafe behavior, and misuse potential, followed by debriefs that convert the experience into articulated professional obligation \cite{borenstein2021}. These exercises serve a dual function, simultaneously training P3 (finding failure) and P4 (owning its implications).

Third, \emph{governance literacy}: working fluency with the risk-management frameworks and statutory regimes that now bind deployed AI systems---including risk classification, documentation, human-oversight, and transparency obligations \cite{nist2023, euaiact2024}. For the agentic engineer, regulation is not compliance overhead but design input: guardrails, audit trails, and oversight mechanisms are engineered artifacts, and their construction belongs in the curriculum alongside every other artifact class \cite{unesco2021, shneiderman2022}.

The thread terminates in the framework's non-negotiable commitment: \textbf{accountability does not delegate}. Whatever autonomy is granted to agents, a named human remains answerable for outcomes. Curricula operationalize this through accountability matrices in projects, sign-off protocols at integration gates (Figure~\ref{fig:loop}), and assessment that asks not only ``does it work?'' but ``who answers for it, and on what evidence?''

\section{Cross-Cutting Thread II --- Assessment: From Artifact to Judgment}
\label{sec:assessment}

Assessment is where the paradigm shift becomes unavoidable. When AI systems solve standard assessments at student level \cite{finnieansley2022}, unassisted-artifact grading loses validity as a measure of professional readiness---and prohibition is neither enforceable nor aligned with the practice graduates will enter \cite{denny2024, prather2023}. Table~\ref{tab:assessment} summarizes the required reorientation: the object of assessment shifts from the artifact to the judgment exercised in producing it.

\begin{table}[t]
\centering
\caption{Reorienting assessment for agentic engineering education.}
\label{tab:assessment}
\small
\begin{tabularx}{\textwidth}{@{}p{3.4cm} L L@{}}
\toprule
\textbf{Dimension} & \textbf{Traditional paradigm} & \textbf{Agentic paradigm} \\
\midrule
Primary object & Final artifact produced without assistance & Judgment: specification quality, delegation rationale, verification rigor, reflective honesty \\
\addlinespace
AI usage & Prohibited or ignored & Declared, documented, and itself assessed \\
\addlinespace
Mode & Summative examination; isolated tasks & Portfolios, design reviews, orchestration logs, defect-detection exercises, viva-style defenses \\
\addlinespace
Integrity model & Detection and prohibition & Transparency and responsible-use demonstration \\
\addlinespace
Foundational knowledge & Assumed measured by artifact production & Measured directly in deliberately AI-restricted components \\
\bottomrule
\end{tabularx}
\end{table}

Three design principles govern the reorientation. First, \emph{assess the loop, not only its output}: orchestration logs, specification documents, review reports, and reflective analyses generated by the delegation--verification loop (Figure~\ref{fig:loop}) are first-class assessment evidence, graded against the pillar outcomes of Table~\ref{tab:pillars}. Process traces are not merely available but diagnostic: a student's prompt-revision trajectory, for example, observably distinguishes reasoning about specification completeness from cosmetic thrashing \cite{lucchetti2025}, giving instructors an evidence-bearing signal of P1 competence that no final artifact can supply. The move toward evaluating reasoning processes, project work, and oral defense---with AI as a declared, legitimate tool---is likewise the convergent recommendation of the broader LLMs-in-education literature \cite{kasneci2023, denny2024}. Second, \emph{protect a foundational core}: because supervision presupposes comprehension, programs retain deliberately AI-restricted assessment components---concept-focused examinations, whiteboard reasoning, live defect-finding---that certify the understanding on which everything else rests \cite{becker2023, sweller2019}. This core is also the countermeasure to the documented ``illusion of competence,'' in which AI-assisted students sincerely overestimate what they have learned \cite{prather2024gap}: self-assessment cannot be trusted to detect the gap that AI-restricted assessment reveals. Third, \emph{make transparency the integrity mechanism}: students declare and document AI usage as professionals must, and the quality of that documentation is itself graded, converting integrity from a policing problem into a competency \cite{prather2023, denny2024}.

AI-supported assessment infrastructure---analytics on collaboration patterns, automated formative feedback on open-ended work---can enrich this regime, provided it is designed under the same learning-sciences discipline demanded of adaptive platforms and is itself subject to the bias and validity scrutiny of Section~\ref{sec:ethics} \cite{luckin2019}. Reducing judgment competencies to convenient proxies would reproduce, inside assessment, exactly the automation credulity the curriculum exists to prevent.

\section{Discussion}
\label{sec:discussion}

\subsection{What ACCEL adds}

ACCEL occupies ground that adjacent frameworks do not. AI-literacy frameworks \cite{long2020, unesco2024students, unesco2024teachers} specify what any citizen, student, or teacher should understand about AI but stop short of the supervisory, orchestration, and accountability competencies specific to professional engineering. Established curriculum guidelines either predate agentic practice entirely \cite{se2014, cc2020} or---as in CS2023, which elevates AI to a core knowledge area and adopts competency-based framing---address AI as subject matter and assistant without yet specifying orchestration, verification, and governance of autonomous collaborators as graduate competencies \cite{cs2023}. Human-factors accounts of automation supervision \cite{lee2004, parasuraman1997, parasuraman2000} diagnose the reliance problem without prescribing an educational architecture for it. And the technical literature on agentic software engineering establishes, comprehensively, what LLM-based agents can do across the lifecycle \cite{liu2026, hou2024}---but leaves open the institutional question of who is educated to direct them, and how; ACCEL is addressed to precisely that residual. Relative to these prior treatments, the framework makes three moves that we regard as its principal contributions. First, it \emph{unifies} literatures that have addressed the problem in isolation: computing-education findings on generative AI \cite{denny2024, prather2023}, human-factors theory on automation supervision \cite{bainbridge1983, lee2004, parasuraman1997, parasuraman2000}, socio-cognitive agency theory and principal--agent delegation theory \cite{bandura2006, jensen1976, baird2021}, and emerging accounts of AI-native practice \cite{alenezi2026delegation, bird2023, barke2023}---yielding an educational architecture in which pedagogical prescriptions inherit theoretical warrant. Second, it \emph{conditions autonomy on verification}: the scaffolded progression (Table~\ref{tab:scaffold}) grants AI autonomy stage-by-stage against demonstrated supervisory competence, directly operationalizing the empirical finding that instructional structure determines whether AI assistance builds or borrows competence \cite{kazemitabaar2023}. Third, it \emph{relocates assessment}: by making the delegation--verification loop itself the object of assessment, the framework restores validity to evaluation in an era when artifacts no longer evidence unassisted competence \cite{finnieansley2022}.

\subsection{Risks the framework is designed to mitigate}

Four failure modes recur in the record, and each is met by a specific framework mechanism. \emph{Automation bias and over-reliance} \cite{parasuraman1997, lee2004}---whose contemporary signature is the measured gap between perceived and actual AI benefit \cite{metr2025, ziegler2024}---are countered by engineered failure encounters, verification-cost design \cite{vasconcelos2023}, and reliance-calibration assessment (Section~\ref{sec:collaboration}). \emph{Deskilling} \cite{bainbridge1983} is countered by the protected foundational core and staged autonomy (Sections~\ref{sec:curricula},~\ref{sec:assessment}). \emph{Superficial engagement}---completing tasks through AI without learning, now documented at scale and known to disproportionately harm weaker students \cite{kazemitabaar2023, prather2024gap, elnaffar2026, borte2023}---is countered by assessing the loop rather than the artifact. \emph{Diffuse accountability} is countered by the non-delegation commitment and its curricular instruments (Section~\ref{sec:ethics}). We emphasize that these are design intentions with empirical warrant, not demonstrated effects; establishing the effects is the first item of the research agenda.

\subsection{Institutional and policy implications}

For universities, the analysis implies that faculty development is the binding constraint: educators cannot teach calibrated supervision of systems they have not themselves learned to supervise, and institutions must resource that learning explicitly \cite{zawacki2019, johri2023, kasneci2023}. The equity stakes deserve emphasis: because AI assistance levels skills on routine work \cite{brynjolfsson2025, peng2023} while effective use presupposes specification and verification competence that novices systematically lack \cite{lucchetti2025, prather2024gap}, unstructured access widens capability gaps even as it appears to democratize practice---scaffolded, explicitly taught progression is therefore an equity instrument, not only a pedagogical preference. Accreditation bodies face pressure to evolve outcome frameworks: even the most recent guidelines, though AI-aware and competency-based \cite{cs2023}, do not yet name orchestration, verification, and governance of autonomous collaborators as explicit graduate outcomes, and older frameworks predate agentic practice entirely \cite{se2014, cc2020}. For governments and national capability programs, ACCEL's continuous-learning vector implies investment in credential architectures and continuing-education taxonomies that keep national workforces adaptive rather than episodically retrained \cite{unesco2021, alenezi2026rise}; national AI academies that integrate degree-level, professional, and executive education within one competency architecture are a natural institutional vehicle.

\subsection{Research agenda}

Five empirical questions follow directly. (1) \emph{Longitudinal skill formation}: do graduates of scaffolded-autonomy programs exhibit stronger unassisted foundations and better-calibrated reliance than graduates of unrestricted-access programs? (2) \emph{Verification pedagogy}: which exercise designs most efficiently produce durable defect-detection skill on AI-generated artifacts, and how does that skill transfer across artifact types? (3) \emph{Assessment validity}: do orchestration-log and portfolio assessments predict professional performance in AI-native teams better than traditional artifact grading? (4) \emph{Trust calibration measurement}: can reliance calibration be instrumented reliably enough---for example, by comparing students' acceptance decisions against measured agent accuracy---to serve as a formal learning outcome? (5) \emph{Institutional change}: which faculty-development and curriculum-governance models allow programs to track a capability frontier that moves faster than revision cycles? Each question is tractable with existing education-research methods; none is answerable from the armchair.

\section{Limitations}
\label{sec:limitations}

Five limitations bound the claims of this article. First, it is a conceptual synthesis: ACCEL is derived from, and consistent with, the empirical record, but the framework itself has not been implemented and evaluated as a whole; its mitigation claims are design hypotheses pending the studies outlined above. Second, the corpus is weighted toward software and computing education, where the empirical record on generative AI is deepest; extension to other engineering disciplines---where physical artifacts, safety certification, and licensure alter the delegation calculus---requires discipline-specific validation. Third, the underlying technology is moving: findings about current model failure modes \cite{vaithilingam2022, dellacqua2023} may age quickly, though we have deliberately anchored the framework in competencies (specification, verification, governance, learning) chosen for robustness to capability shifts rather than in tool-specific skills. Fourth, the analysis largely presumes institutional contexts with the resources to execute systemic reform; adaptation to resource-constrained settings, where faculty capacity and infrastructure are limiting, is an open design problem of the first importance \cite{unesco2021}. Fifth, the synthesis was conducted by a single author: corpus coding was not subject to independent inter-rater checks, and a portion of the sources characterizing AI-native practice are the author's own recent work, some of it available only as preprints at the time of writing. We have mitigated both risks by the cross-community convergence rule of Section~\ref{sec:method} and by anchoring every load-bearing claim to at least one independent, peer-reviewed source, but readers should weigh these provenance facts when assessing the argument.

\section{Conclusion}
\label{sec:conclusion}

The rise of agentic AI does not diminish the engineer; it relocates the engineer's value. When autonomous systems can produce artifacts, the scarce and durable human contributions become the framing of intent, the orchestration of hybrid teams, the verification of machine work, the ownership of ethical consequence, and the discipline of continuous adaptation. Educating for these contributions is not accomplished by adding a course, licensing a tool, or prohibiting one. It requires the systemic re-architecture this article has specified: competency pillars grounded in agency theory and human-factors evidence; curricula that grant AI autonomy only as verification competence is demonstrated; collaboration structures that make the delegation--verification loop explicit, teachable, and assessable; assessment that measures judgment rather than artifacts; ethics practiced as governance engineering rather than recited as code; and learning designed to outlast the degree that begins it.

The stakes exceed pedagogy. Societies are delegating consequential decisions to increasingly autonomous systems, and the engineers who supervise those systems constitute the profession's---and the public's---primary line of accountable oversight. Whether that oversight is exercised with calibrated judgment or credulous deference will be decided, in large part, by how the current generation of engineers is educated. The agentic engineer, in the full Bandurian sense, is therefore this era's central educational objective: a professional whose learning never stops, whose collaboration spans human and artificial minds, and whose agency ensures that autonomous technology remains an instrument of human intention, dignity, and progress.

\bibliographystyle{unsrt}
\bibliography{references}

@article{alenezi2026delegation,
  author  = {Alenezi, Mamdouh},
  title   = {From Determinism to Delegation: {AI}-Native Software Engineering and the Evolution of the Agentic Engineer},
  journal = {arXiv preprint arXiv:2606.28791},
  year    = {2026}
}

@article{alenezi2026rise,
  author  = {Alenezi, Mamdouh},
  title   = {The Rise of {AI}-Native Software Engineering: Implications for Practice, Education, and the Future Workforce},
  journal = {arXiv preprint arXiv:2606.12986},
  year    = {2026}
}

@article{alenezi2026collaboration,
  author  = {Alenezi, Mamdouh},
  title   = {Human--{AI} Collaboration and the Transformation of Software Engineering Work},
  journal = {arXiv preprint arXiv:2606.03394},
  year    = {2026}
}

@article{bandura2006,
  author  = {Bandura, Albert},
  title   = {Toward a Psychology of Human Agency},
  journal = {Perspectives on Psychological Science},
  volume  = {1},
  number  = {2},
  pages   = {164--180},
  year    = {2006}
}

@inproceedings{long2020,
  author    = {Long, Duri and Magerko, Brian},
  title     = {What is {AI} Literacy? Competencies and Design Considerations},
  booktitle = {Proceedings of the 2020 CHI Conference on Human Factors in Computing Systems},
  pages     = {1--16},
  year      = {2020},
  publisher = {ACM}
}

@article{zawacki2019,
  author  = {Zawacki-Richter, Olaf and Mar{\'i}n, Victoria I. and Bond, Melissa and Gouverneur, Franziska},
  title   = {Systematic Review of Research on Artificial Intelligence Applications in Higher Education---Where Are the Educators?},
  journal = {International Journal of Educational Technology in Higher Education},
  volume  = {16},
  number  = {1},
  pages   = {39},
  year    = {2019}
}

@article{borte2023,
  author  = {B{\o}rte, Kristin and Nesje, Katrine and Lillejord, S{\o}lvi},
  title   = {Barriers to Student Active Learning in Higher Education},
  journal = {Teaching in Higher Education},
  volume  = {28},
  number  = {3},
  pages   = {597--615},
  year    = {2023}
}

@article{froyd2012,
  author  = {Froyd, Jeffrey E. and Wankat, Phillip C. and Smith, Karl A.},
  title   = {Five Major Shifts in 100 Years of Engineering Education},
  journal = {Proceedings of the IEEE},
  volume  = {100},
  pages   = {1344--1360},
  year    = {2012}
}

@article{lattuca2017,
  author  = {Lattuca, Lisa R. and Knight, David B. and Ro, Hyun Kyoung and Novoselich, Brian J.},
  title   = {Supporting the Development of Engineers' Interdisciplinary Competence},
  journal = {Journal of Engineering Education},
  volume  = {106},
  number  = {1},
  pages   = {71--97},
  year    = {2017}
}

@article{sengul2024,
  author  = {Sengul, Cigdem and Neykova, Rumyana and Destefanis, Giuseppe},
  title   = {Software Engineering Education in the Era of Conversational {AI}: Current Trends and Future Directions},
  journal = {Frontiers in Artificial Intelligence},
  volume  = {7},
  pages   = {1436350},
  year    = {2024}
}

@article{johri2023,
  author  = {Johri, Aditya and Katz, Andrew S. and Qadir, Junaid and Kohli, Ashish},
  title   = {Generative Artificial Intelligence and Engineering Education},
  journal = {Journal of Engineering Education},
  volume  = {112},
  number  = {3},
  pages   = {572--577},
  year    = {2023}
}

@article{memarian2023,
  author  = {Memarian, Bahar and Doleck, Tenzin},
  title   = {Fairness, Accountability, Transparency, and Ethics ({FATE}) in Artificial Intelligence ({AI}) and Higher Education: A Systematic Review},
  journal = {Computers and Education: Artificial Intelligence},
  volume  = {5},
  pages   = {100152},
  year    = {2023}
}

@article{denny2024,
  author  = {Denny, Paul and Prather, James and Becker, Brett A. and Finnie-Ansley, James and Hellas, Arto and Leinonen, Juho and Luxton-Reilly, Andrew and Reeves, Brent N. and Santos, Eddie Antonio and Sarsa, Sami},
  title   = {Computing Education in the Era of Generative {AI}},
  journal = {Communications of the ACM},
  volume  = {67},
  number  = {2},
  pages   = {56--67},
  year    = {2024}
}

@inproceedings{prather2023,
  author    = {Prather, James and Denny, Paul and Leinonen, Juho and Becker, Brett A. and Albluwi, Ibrahim and Craig, Michelle and Keuning, Hieke and Kiesler, Natalie and Kohn, Tobias and Luxton-Reilly, Andrew and MacNeil, Stephen and Petersen, Andrew and Pettit, Raymond and Reeves, Brent N. and Savelka, Jaromir},
  title     = {The Robots Are Here: Navigating the Generative {AI} Revolution in Computing Education},
  booktitle = {Proceedings of the 2023 Working Group Reports on Innovation and Technology in Computer Science Education (ITiCSE-WGR '23)},
  pages     = {108--159},
  year      = {2023},
  publisher = {ACM}
}

@inproceedings{finnieansley2022,
  author    = {Finnie-Ansley, James and Denny, Paul and Becker, Brett A. and Luxton-Reilly, Andrew and Prather, James},
  title     = {The Robots Are Coming: Exploring the Implications of {OpenAI} {Codex} on Introductory Programming},
  booktitle = {Proceedings of the 24th Australasian Computing Education Conference (ACE '22)},
  pages     = {10--19},
  year      = {2022},
  publisher = {ACM}
}

@inproceedings{kazemitabaar2023,
  author    = {Kazemitabaar, Majeed and Chow, Justin and Ma, Carl Ka To and Ericson, Barbara J. and Weintrop, David and Grossman, Tovi},
  title     = {Studying the Effect of {AI} Code Generators on Supporting Novice Learners in Introductory Programming},
  booktitle = {Proceedings of the 2023 CHI Conference on Human Factors in Computing Systems},
  pages     = {1--23},
  year      = {2023},
  publisher = {ACM}
}

@inproceedings{becker2023,
  author    = {Becker, Brett A. and Denny, Paul and Finnie-Ansley, James and Luxton-Reilly, Andrew and Prather, James and Santos, Eddie Antonio},
  title     = {Programming Is Hard---Or at Least It Used to Be: Educational Opportunities and Challenges of {AI} Code Generation},
  booktitle = {Proceedings of the 54th ACM Technical Symposium on Computer Science Education (SIGCSE '23)},
  pages     = {500--506},
  year      = {2023},
  publisher = {ACM}
}

@article{peng2023,
  author  = {Peng, Sida and Kalliamvakou, Eirini and Cihon, Peter and Demirer, Mert},
  title   = {The Impact of {AI} on Developer Productivity: Evidence from {GitHub} {Copilot}},
  journal = {arXiv preprint arXiv:2302.06590},
  year    = {2023}
}

@article{bird2023,
  author  = {Bird, Christian and Ford, Denae and Zimmermann, Thomas and Forsgren, Nicole and Kalliamvakou, Eirini and Lowdermilk, Travis and Gazit, Idan},
  title   = {Taking Flight with {Copilot}},
  journal = {Communications of the ACM},
  volume  = {66},
  number  = {6},
  pages   = {56--62},
  year    = {2023}
}

@inproceedings{vaithilingam2022,
  author    = {Vaithilingam, Priyan and Zhang, Tianyi and Glassman, Elena L.},
  title     = {Expectation vs. Experience: Evaluating the Usability of Code Generation Tools Powered by Large Language Models},
  booktitle = {Extended Abstracts of the 2022 CHI Conference on Human Factors in Computing Systems},
  pages     = {1--7},
  year      = {2022},
  publisher = {ACM}
}

@inproceedings{sarkar2022,
  author    = {Sarkar, Advait and Gordon, Andrew D. and Negreanu, Carina and Poelitz, Christian and Ragavan, Sruti Srinivasa and Zorn, Ben},
  title     = {What Is It Like to Program with Artificial Intelligence?},
  booktitle = {Proceedings of the 33rd Annual Conference of the Psychology of Programming Interest Group (PPIG 2022)},
  year      = {2022}
}

@inproceedings{amershi2019,
  author    = {Amershi, Saleema and Weld, Dan and Vorvoreanu, Mihaela and Fourney, Adam and Nushi, Besmira and Collisson, Penny and Suh, Jina and Iqbal, Shamsi and Bennett, Paul N. and Inkpen, Kori and Teevan, Jaime and Kikin-Gil, Ruth and Horvitz, Eric},
  title     = {Guidelines for Human--{AI} Interaction},
  booktitle = {Proceedings of the 2019 CHI Conference on Human Factors in Computing Systems},
  pages     = {1--13},
  year      = {2019},
  publisher = {ACM}
}

@article{shneiderman2020,
  author  = {Shneiderman, Ben},
  title   = {Human-Centered Artificial Intelligence: Reliable, Safe \& Trustworthy},
  journal = {International Journal of Human--Computer Interaction},
  volume  = {36},
  number  = {6},
  pages   = {495--504},
  year    = {2020}
}

@article{lee2004,
  author  = {Lee, John D. and See, Katrina A.},
  title   = {Trust in Automation: Designing for Appropriate Reliance},
  journal = {Human Factors},
  volume  = {46},
  number  = {1},
  pages   = {50--80},
  year    = {2004}
}

@article{parasuraman1997,
  author  = {Parasuraman, Raja and Riley, Victor},
  title   = {Humans and Automation: Use, Misuse, Disuse, Abuse},
  journal = {Human Factors},
  volume  = {39},
  number  = {2},
  pages   = {230--253},
  year    = {1997}
}

@article{bainbridge1983,
  author  = {Bainbridge, Lisanne},
  title   = {Ironies of Automation},
  journal = {Automatica},
  volume  = {19},
  number  = {6},
  pages   = {775--779},
  year    = {1983}
}

@techreport{dellacqua2023,
  author      = {Dell'Acqua, Fabrizio and McFowland III, Edward and Mollick, Ethan R. and Lifshitz-Assaf, Hila and Kellogg, Katherine and Rajendran, Saran and Krayer, Lisa and Candelon, Fran{\c{c}}ois and Lakhani, Karim R.},
  title       = {Navigating the Jagged Technological Frontier: Field Experimental Evidence of the Effects of {AI} on Knowledge Worker Productivity and Quality},
  institution = {Harvard Business School},
  number      = {Working Paper 24-013},
  year        = {2023}
}

@article{borenstein2021,
  author  = {Borenstein, Jason and Howard, Ayanna},
  title   = {Emerging Challenges in {AI} and the Need for {AI} Ethics Education},
  journal = {AI and Ethics},
  volume  = {1},
  pages   = {61--65},
  year    = {2021}
}

@book{unesco2021,
  author    = {Miao, Fengchun and Holmes, Wayne and Huang, Ronghuai and Zhang, Hui},
  title     = {{AI} and Education: Guidance for Policy-Makers},
  publisher = {UNESCO},
  address   = {Paris},
  year      = {2021}
}

@techreport{se2014,
  author      = {{ACM/IEEE-CS Joint Task Force on Computing Curricula}},
  title       = {Software Engineering 2014: Curriculum Guidelines for Undergraduate Degree Programs in Software Engineering},
  institution = {ACM and IEEE Computer Society},
  year        = {2015}
}

@techreport{cc2020,
  author      = {{CC2020 Task Force}},
  title       = {Computing Curricula 2020: Paradigms for Global Computing Education},
  institution = {ACM and IEEE Computer Society},
  year        = {2020}
}

@article{luckin2019,
  author  = {Luckin, Rose and Cukurova, Mutlu},
  title   = {Designing Educational Technologies in the Age of {AI}: A Learning Sciences-Driven Approach},
  journal = {British Journal of Educational Technology},
  volume  = {50},
  number  = {6},
  pages   = {2824--2838},
  year    = {2019}
}

@article{sweller2019,
  author  = {Sweller, John and van Merri{\"e}nboer, Jeroen J. G. and Paas, Fred},
  title   = {Cognitive Architecture and Instructional Design: 20 Years Later},
  journal = {Educational Psychology Review},
  volume  = {31},
  pages   = {261--292},
  year    = {2019}
}

@article{chen2021,
  author  = {Chen, Juebei and Kolmos, Anette and Du, Xiangyun},
  title   = {Forms of Implementation and Challenges of {PBL} in Engineering Education: A Review of Literature},
  journal = {European Journal of Engineering Education},
  volume  = {46},
  number  = {1},
  pages   = {90--115},
  year    = {2021}
}

@article{fischer2000,
  author  = {Fischer, Gerhard},
  title   = {Lifelong Learning---More Than Training},
  journal = {Journal of Interactive Learning Research},
  volume  = {11},
  number  = {3},
  pages   = {265--294},
  year    = {2000}
}

@book{schon1983,
  author    = {Sch{\"o}n, Donald A.},
  title     = {The Reflective Practitioner: How Professionals Think in Action},
  publisher = {Basic Books},
  address   = {New York},
  year      = {1983}
}

@article{zimmerman2002,
  author  = {Zimmerman, Barry J.},
  title   = {Becoming a Self-Regulated Learner: An Overview},
  journal = {Theory Into Practice},
  volume  = {41},
  number  = {2},
  pages   = {64--70},
  year    = {2002}
}

@techreport{nist2023,
  author      = {{National Institute of Standards and Technology}},
  title       = {Artificial Intelligence Risk Management Framework ({AI} {RMF} 1.0)},
  institution = {U.S. Department of Commerce},
  number      = {NIST AI 100-1},
  year        = {2023}
}

@misc{euaiact2024,
  author       = {{European Parliament and Council of the European Union}},
  title        = {Regulation ({EU}) 2024/1689 Laying Down Harmonised Rules on Artificial Intelligence (Artificial Intelligence Act)},
  howpublished = {Official Journal of the European Union, L series},
  year         = {2024}
}

@article{akour2024,
  author  = {Akour, Mohammed and Alenezi, Mamdouh},
  title   = {The Enduring Impact of Gamification on Software Engineering Students' Engagement},
  journal = {International Journal of Technology Enhanced Learning},
  year    = {2024}
}

@inproceedings{akour2025,
  author    = {Akour, Mohammed and Alenezi, Mamdouh},
  title     = {Agentic {AI} in {DevOps}: Boosting Software Automation and Collaboration},
  booktitle = {2025 International Conference on Artificial Intelligence Security and Applications},
  year      = {2025},
  publisher = {IEEE}
}

@inproceedings{prather2024gap,
  author    = {Prather, James and Reeves, Brent N. and Leinonen, Juho and MacNeil, Stephen and Randrianasolo, Arisoa S. and Becker, Brett A. and Kimmel, Bailey and Wright, Jared and Briggs, Ben},
  title     = {The Widening Gap: The Benefits and Harms of Generative {AI} for Novice Programmers},
  booktitle = {Proceedings of the 2024 ACM Conference on International Computing Education Research (ICER '24)},
  pages     = {469--486},
  year      = {2024},
  publisher = {ACM}
}

@techreport{cs2023,
  author      = {{ACM/IEEE-CS/AAAI Joint Task Force on Computing Curricula}},
  title       = {Computer Science Curricula 2023 ({CS2023}): The Final Report},
  institution = {ACM, IEEE Computer Society, and AAAI},
  year        = {2024}
}

@book{unesco2024students,
  author    = {Miao, Fengchun and Shiohira, Kelly},
  title     = {{AI} Competency Framework for Students},
  publisher = {UNESCO},
  address   = {Paris},
  year      = {2024}
}

@book{unesco2024teachers,
  author    = {Miao, Fengchun and Cukurova, Mutlu},
  title     = {{AI} Competency Framework for Teachers},
  publisher = {UNESCO},
  address   = {Paris},
  year      = {2024}
}

@article{metr2025,
  author  = {Becker, Joel and Rush, Nate and Barnes, Beth and Rein, David},
  title   = {Measuring the Impact of Early-2025 {AI} on Experienced Open-Source Developer Productivity},
  journal = {arXiv preprint arXiv:2507.09089},
  year    = {2025}
}

@article{elnaffar2026,
  author  = {Elnaffar, Said and Rashidi, Farzad and Abualkishik, Abedallah Zaid},
  title   = {Teaching with {AI}: A Systematic Review of Chatbots, Generative Tools, and Tutoring Systems in Programming Education},
  journal = {International Journal of Learning, Teaching and Educational Research},
  volume  = {25},
  number  = {1},
  pages   = {1--28},
  year    = {2026}
}

@article{snyder2019,
  author  = {Snyder, Hannah},
  title   = {Literature Review as a Research Methodology: An Overview and Guidelines},
  journal = {Journal of Business Research},
  volume  = {104},
  pages   = {333--339},
  year    = {2019}
}

@article{torraco2005,
  author  = {Torraco, Richard J.},
  title   = {Writing Integrative Literature Reviews: Guidelines and Examples},
  journal = {Human Resource Development Review},
  volume  = {4},
  number  = {3},
  pages   = {356--367},
  year    = {2005}
}

@article{baird2021,
  author  = {Baird, Aaron and Maruping, Likoebe M.},
  title   = {The Next Generation of Research on {IS} Use: A Theoretical Framework of Delegation to and from Agentic {IS} Artifacts},
  journal = {MIS Quarterly},
  volume  = {45},
  number  = {1},
  pages   = {315--341},
  year    = {2021}
}

@article{jensen1976,
  author  = {Jensen, Michael C. and Meckling, William H.},
  title   = {Theory of the Firm: Managerial Behavior, Agency Costs and Ownership Structure},
  journal = {Journal of Financial Economics},
  volume  = {3},
  number  = {4},
  pages   = {305--360},
  year    = {1976}
}

@article{brynjolfsson2025,
  author  = {Brynjolfsson, Erik and Li, Danielle and Raymond, Lindsey},
  title   = {Generative {AI} at Work},
  journal = {The Quarterly Journal of Economics},
  volume  = {140},
  number  = {2},
  pages   = {889--942},
  year    = {2025}
}

@article{ziegler2024,
  author  = {Ziegler, Albert and Kalliamvakou, Eirini and Li, X. Alice and Rice, Andrew and Rifkin, Devon and Simister, Shawn and Sittampalam, Ganesh and Aftandilian, Edward},
  title   = {Measuring {GitHub} {Copilot}'s Impact on Productivity},
  journal = {Communications of the ACM},
  volume  = {67},
  number  = {3},
  pages   = {54--61},
  year    = {2024}
}

@article{vasconcelos2023,
  author  = {Vasconcelos, Helena and J{\"o}rke, Matthew and Grunde-McLaughlin, Madeleine and Gerstenberg, Tobias and Bernstein, Michael S. and Krishna, Ranjay},
  title   = {Explanations Can Reduce Overreliance on {AI} Systems During Decision-Making},
  journal = {Proceedings of the ACM on Human-Computer Interaction},
  volume  = {7},
  number  = {CSCW1},
  pages   = {Article 129},
  year    = {2023}
}

@inproceedings{park2023,
  author    = {Park, Joon Sung and O'Brien, Joseph C. and Cai, Carrie J. and Morris, Meredith Ringel and Liang, Percy and Bernstein, Michael S.},
  title     = {Generative Agents: Interactive Simulacra of Human Behavior},
  booktitle = {Proceedings of the 36th Annual ACM Symposium on User Interface Software and Technology (UIST '23)},
  pages     = {1--22},
  year      = {2023},
  publisher = {ACM}
}

@article{hou2024,
  author  = {Hou, Xinyi and Zhao, Yanjie and Liu, Yue and Yang, Zhou and Wang, Kailong and Li, Li and Luo, Xiapu and Lo, David and Grundy, John and Wang, Haoyu},
  title   = {Large Language Models for Software Engineering: A Systematic Literature Review},
  journal = {ACM Transactions on Software Engineering and Methodology},
  volume  = {33},
  number  = {8},
  pages   = {Article 220},
  year    = {2024}
}

@article{mittelstadt2019,
  author  = {Mittelstadt, Brent},
  title   = {Principles Alone Cannot Guarantee Ethical {AI}},
  journal = {Nature Machine Intelligence},
  volume  = {1},
  number  = {11},
  pages   = {501--507},
  year    = {2019}
}

@article{kasneci2023,
  author  = {Kasneci, Enkelejda and Se{\ss}ler, Kathrin and K{\"u}chemann, Stefan and Bannert, Maria and Dementieva, Daryna and Fischer, Frank and Gasser, Urs and Groh, Georg and G{\"u}nnemann, Stephan and H{\"u}llermeier, Eyke and Krusche, Stephan and Kutyniok, Gitta and Michaeli, Tilman and Nerdel, Claudia and Pfeffer, J{\"u}rgen and Poquet, Oleksandra and Sailer, Michael and Schmidt, Albrecht and Seidel, Tina and Stadler, Matthias and Weller, Jochen and Kuhn, Jochen and Kasneci, Gjergji},
  title   = {{ChatGPT} for Good? On Opportunities and Challenges of Large Language Models for Education},
  journal = {Learning and Individual Differences},
  volume  = {103},
  pages   = {102274},
  year    = {2023}
}

@article{parasuraman2000,
  author  = {Parasuraman, Raja and Sheridan, Thomas B. and Wickens, Christopher D.},
  title   = {A Model for Types and Levels of Human Interaction with Automation},
  journal = {IEEE Transactions on Systems, Man, and Cybernetics---Part A: Systems and Humans},
  volume  = {30},
  number  = {3},
  pages   = {286--297},
  year    = {2000}
}

@book{shneiderman2022,
  author    = {Shneiderman, Ben},
  title     = {Human-Centered {AI}},
  publisher = {Oxford University Press},
  address   = {Oxford},
  year      = {2022}
}

@book{nasem2018,
  author    = {{National Academies of Sciences, Engineering, and Medicine}},
  title     = {How People Learn {II}: Learners, Contexts, and Cultures},
  publisher = {The National Academies Press},
  address   = {Washington, DC},
  year      = {2018}
}

@article{barke2023,
  author  = {Barke, Shraddha and James, Michael B. and Polikarpova, Nadia},
  title   = {Grounded {Copilot}: How Programmers Interact with Code-Generating Models},
  journal = {Proceedings of the ACM on Programming Languages},
  volume  = {7},
  number  = {OOPSLA1},
  pages   = {85--111},
  year    = {2023}
}

@article{liu2026,
  author  = {Liu, Junwei and Wang, Kaixin and Chen, Yixuan and Peng, Xin and Chen, Zhenpeng and Zhang, Lingming and Lou, Yiling},
  title   = {Large Language Model-Based Agents for Software Engineering: A Survey},
  journal = {ACM Transactions on Software Engineering and Methodology},
  year    = {2026},
  note    = {\url{https://doi.org/10.1145/3796507}}
}

@inproceedings{lucchetti2025,
  author    = {Lucchetti, Francesca and Wu, Zixuan and Guha, Arjun and Feldman, Molly Q. and Anderson, Carolyn Jane},
  title     = {Substance Beats Style: Why Beginning Students Fail to Code with {LLMs}},
  booktitle = {Proceedings of the 2025 Conference of the Nations of the Americas Chapter of the Association for Computational Linguistics (NAACL-HLT 2025), Volume 1: Long Papers},
  pages     = {8541--8610},
  year      = {2025},
  publisher = {Association for Computational Linguistics}
}

\end{document}